# Crystal Growth & Physical Property Characterization of Mixed Topological Insulator BiSbTe$_3$

Dinesh Kumar[1], Kapil Kumar[2,3], N. K. Karn[2,3], Ganesh Gurjar[4], V.P.S. Awana[2,3,*], and Sudesh[1]

*[1]Department of Physical Sciences, Banasthali Vidyapith, Newai-304022, India*
*[2]CSIR-National Physical Laboratory, Dr. K.S. Krishnan Marg, New Delhi-110012, India*
*[3]Academy of Scientific and Innovative Research (AcSIR), Ghaziabad-201002, India*
*[4]Department of Physics, Ramjas College, University of Delhi, Delhi-110007, India*

This article reports the synthesis of a single crystalline mixed topological insulator (TI) BiSbTe$_3$ and its detailed structural and magneto-transport properties. The single crystalline samples of BiSbTe$_3$ are grown by the melt-growth process and characterized by X-ray diffraction (XRD), Energy dispersive X-ray analysis (EDAX) and Raman spectroscopy. The single crystal XRD peaks dictated the growth direction along the c-axis. The Raman spectrum elucidated the characteristic peaks of the mixed topological insulator. The broadening of Raman peaks exhibited the formation of Te-Bi-Te and Te-Sb-Te bonds and associated vibrational modes. The single crystals are characterized by magneto-transport measurements down to 2 K and up to 14 Tesla transverse magnetic field. The residual resistance ratio ($R_{200\,K}/R_{0\,K}$) is found to be 3.64, which endorses the metallic nature of the synthesized crystal. The relative resistance turns out to be higher for the mixed TI than the pure TIs i.e., Bi$_2$Te$_3$ or Sb$_2$Te$_3$. The lower Debye temperature (82.64 K) of BiSbTe$_3$ connotes the presence of effective electron-phonon interaction at quite low temperatures in comparison to pure TI, which explains the observed suppression in magnetoresistance (MR) for the mixed TI. At 2 K, an MR of ~150% is observed for BiSbTe$_3$, which is suppressed in contrast to the pure TIs i.e., Bi$_2$Te$_3$ or Sb$_2$Te$_3$. Though the MR% is suppressed significantly, its non-saturating linear behavior indicates the topological nature of the studied mixed TI. The modified Hikami-Larkin-Nagaoka (HLN) equation analysis of magneto-conductivity of mixed TI revealed that the conductivity has not only a surface states driven 2D component but also contributions from the bulk charge carriers and quantum scattering.

**Keywords**: Crystal Growth, Topological Insulator, Magnetoresistance, HLN analysis

*Corresponding Author
Dr. V.P.S. Awana: Chief Scientist,
CSIR-NPL, New Delhi,
India, 110012
E-mail: awana@nplindia.res.in
Ph. +91-11-45609357
Fax-+91-11-45609310

**Introduction**

Topological insulators (TIs) have gained significant attention recently due to their remarkable properties. The TIs are novel quantum materials with an insulating bulk and robust conducting surface states (SSs) [1-3]. The robustness of the SSs originates from the time-reversal symmetry (TRS) [4]. Moreover, the TRS leads to intrinsic spin-orbit coupling (SOC) in TIs [5, 6], which discards possibility of backscattering by non-magnetic impurities and give rise to spin polarization in these materials. The numbers of SSs are odd in these materials and are regarded as massless Dirac Fermion states [7]. The topologically protected conducting SSs are present between the bulk conduction band (BCB), and bulk valance band (BVB) and can be experimentally seen in angle-resolved photoelectron spectroscopy (ARPES) spectra in the form of two inverted cones, which meet each other at the Dirac point [8]. The insulating bulk in 3D TIs allows the dissipation-less flow of charge carriers through their conducting surface states. The unique properties of the conducting SSs in topological insulators make these materials potentially valuable for applications in the field of spintronics as well as in quantum computing [9-12]. Moreover, the strong spin-orbit coupling (SOC) and the resulting high MR facilitate the development of more reliable magnetic memory devices and magnetic sensors [13]. Further TIs, also enable the observation of quantum phenomena such as Shubnikov-de Haas oscillations [14, 15], and rare oscillations like log-periodic quantum oscillations [16].

Some binary tetradymite compounds, including $Bi_2Se_3$, $Bi_2Te_3$, and $Sb_2Te_3$, have emerged as model quantum topological materials and have been the focus of extensive research in recent years [1-17]. The topological nature has been experimentally confirmed through the presence of Dirac cone in various TIs such as $Bi_2Se_3$ and $Bi_2Te_3$ using ARPES [1, 5]. Studies have shown that, compared to conventional insulators, these materials exhibit a relatively small band gap (~0.3–0.4 eV). Consequently, some degree of conductivity is observed at room temperature, indicating that the bulk is not fully insulating but rather semiconducting. However, in the low-temperature regime, they resemble a conductor instead of an insulator due to the presence of surface states. To improve the bulk insulation properties, ternary tetradymites or mixed TIs, such as $Bi_2Te_2Se$ and $Bi_2TeSe_2$, have been developed [17-19]. These mixed TI materials retain a layered crystal structure similar to that of the tetradymite family [19].

Electromagnetic responses in magneto-transport measurements, such as non-saturating linear MR [20], negative MR [21], and oscillations in MR (SdH)[22], serve as indirect probes for the surface states. Theoretically, topological surface states are analyzed and identified using associated invariants, such as the Z2 invariant [23]. The analysis of magneto-conductivity using the modified Hikami-Larkin-Nagaoka (HLN) equation [24, 25] indicates that conductivity comprises not only a two-dimensional surface component but also contributions from bulk charge carriers and quantum scattering effects.

Among various possible mixed TIs, $BiSbTe_3$ have been widely reported, and its non-trivial topological character with Dirac like charge carriers has been revealed by ARPES measurements [26] and DFT calculations [27]. The quantum oscillations in magnetoresistance measurements with $\pi$ Berry phase confirm the Dirac charge carriers present in the mixed TI $BiSbTe_3$ [28]. However, in previous reports the role of phonons is not well identified and discussed as per our knowledge. Also, the work by Rajput et al [29], reports magneto-conductivity analysis but the stoichiometry of the crystals is far from the ideal. Therefore, it becomes essential to study the Mixed TI $BiSbTe_3$ covering these aspects with better stoichiometric compositions.

The growth of single crystals of this class of compounds, typically requires 8–10 days for synthesis. In this study, a similar method has been adopted, but with shorter heating and cooling durations leading to faster growth of single crystal of $BiSbTe_3$. The resulting crystals have been evaluated for their structural and electrical properties, with XRD, EDAX, Scanning Electron Microscopy(SEM), and Raman analysis to confirm their good quality. Topological behavior is being identified through temperature-dependent and field-dependent resistivity measurements. Resistance-temperature (RT) measurements were conducted down to 2 K, while MR was measured in a transverse magnetic field range of ±14 Tesla. The HLN analysis of magneto-conductivity has been performed to decode the surface, bulk and other scattering contributions present in the as-grown single crystalline mixed TI $BiSbTe_3$.

**Experimental**

There are several methods to grow single crystals depending on the type of material and crystallite size required. Among them common methods include melt-growth [30] and the vertical Bridgman technique [31] for both inorganic and organic crystals and solution growth for organic crystals [32]. Here we have used the melt-growth technique to grow the single crystals of $BiSbTe_3$. High purity (4N) powders of Bi (Sigma Aldrich 99.9%), Sb(Sigma Aldrich 99.5%) and Te (Glasil 99.9%) were used in their stoichiometric ratio Bi:Sb:Te as 1:1:3 and mixed thoroughly using the agate-mortar and pestle. The mixture is then palletized in a hydraulic press palletizer and vacuum-sealed in a quartz tube at a pressure of $5×10^{-4}$ mbar. The sealed pallets are then subjected to heat treatment following the scheme as shown in Fig. 1. The vacuum sealed pallet was first heated up to 900° C at a rate of 60° C/h and kept at this elevated temperature for 48 hours. Subsequently, this homogenous melt was allowed to cool down to room temperature at a rate of 60° C/h. Thus, obtained single crystals are silvery shiny and easily cleavable along their growth direction. The image of the as-grown $BiSbTe_3$ single crystal is shown in the inset of Fig. 1.

The as-grown samples are characterized for their phase purity and chemical composition. Rigaku made Mini flex II X-ray diffractometer with Cu Kα radiation of 1.5418 Å wavelength is used to record X-ray Diffraction (XRD) pattern on both mechanically cleaved single crystalline flakes and gently crushed powder of synthesized crystals at room temperature. For the XRD measurement the scan rate is set to $2^0$/min and step size is $0.02^0$ with 30 kV supply and 15 mA current output to X-ray tube. The powder XRD pattern is fitted using FullProf Software to analyze X-ray diffraction (XRD) data, particularly through Rietveld refinement. Raman spectra of synthesized crystals are recorded by using Renishaw in Via Reflex Raman Microscope equipped with Laser of 514 nm. The focal length of the spectrometer is 250 mm and the grating density is 2400 l/mm. The absolute accuracy of the spectrometer is around ±1 $cm^{-1}$. Mechanically cleaved flakes of synthesized crystals are irradiated with a Laser of wavelength 514 nm for 30 sec. During the irradiation process, the Laser power is maintained below 5 mW to avoid any local heating on the crystal surface. EDAX and SEM has been performed to corroborate the elemental composition and probe the surface morphology and layered structure of as-synthesized crystal, respectively, using the SIGMA VP model High-Resolution Field Emission Scanning Electron Microscope (HR FESEM). The magneto-transport studies were performed via a conventional four-probe method on a Quantum Design Physical Property Measurement System (PPMS) equipped with a sample rotator and closed-cycle based cryogen-free system. For the magneto-transport measurements, the current was supplied in ab-plane of the crystal, while the magnetic field was kept perpendicular to the ab-plane.

**Results and Discussion**

The fresh silvery shiny chunk of the BiSbTe$_3$ is obtained, as shown in the inset of Fig. 1, out of which single crystals are extracted by cleaving the chunk mechanically. The XRD pattern is recorded on the mechanically cleaved single crystal flakes of BiSbTe$_3$ as shown in Fig. 2(a). These peaks are identified as (003n) in accordance with previous reports[29], which exhibits the unidirectional growth along c-axis of the as-grown BiSbTe$_3$ crystal. The full width at half maximum (FWHM) for the (0015) (highest intensity peak) is found to be 0.14°. This shows the highly crystalline nature of the grown BiSbTe$_3$ along the c-axis. The lattice parameter c obtained from the single crystal XRD peaks is 30.455(4) Å, which is comparable to the previous report[29, 33] and agrees with the Rietveld refinement of powder XRD patterns. Fig. 2(b) depicts the powdered XRD data and its Rietveld refinement results for the as-grown crystal. The mixed TI BiSbTe$_3$ has rhombohedral structure, isostructural to the parent TI compounds Sb$_2$Te$_3$ and Bi$_2$Te$_3$. To confirm the phase purity of the as-grown BiSbTe$_3$, the powder XRD data is refined by Rietveld method implemented in FullProf software. It affirms that the synthesized crystal has rhombohedral lattice with (R-3m) space group. The quality of fit is estimated by $\chi^2$ (parameter for the goodness of fit), which is estimated to be 2.89, which is in an acceptable range. The refined lattice parameters are $a = b = 4.329(4)$ Å, $c = 30.438(2)$ Å, $\alpha = \beta = 90°$ and $\gamma = 120°$. The Rietveld refinement results are summarized in Table 1. Using the refined lattice parameters, VESTA generated unit cell is depicted in Fig. 2(c) exhibiting the stacking of quintuple layers.

Fig. 3(a) displays an FESEM images of synthesized single-crystalline BiSbTe$_3$ at 5 μm resolution, revealing a layered terrace morphology as straight line-type color contrast showing cleavable edges present in the grown sample. This confirms single-crystalline growth along the c-axis. Fig. 3(b) depicts another FESEM image taken on a single plane of the cleaved flake at 20 μm resolution. The absence of color contrast indicates a single-phase structure. To further confirm the compositional uniformity and purity of the grown crystal, elemental mapping has been performed. Fig. 3(c)-(e) depicts the elemental mapping corresponding to Bi, Sb and Te elements respectively. All the compositional elements are distributed uniformly throughout on the measured sample surface. Fig. 3(f) shows EDAX spectra confirming the exclusive presence of Bi, Sb and Te, with no impurities. The inset table shows the weight percentage and atomic percentage of the constituent elements resulting in the stoichiometry Bi$_{0.98}$ Sb$_{1.04}$Te$_{2.97}$, which aligns with BiSbTe$_3$ within the experimental error.

Vibrational modes have been studied by the Raman spectroscopy. The binary tetradymite compounds have four Raman active modes, namely $E_g^1$, $E_g^2$, $A_{1g}^1$ and $A_{1g}^2$ [34]. Out of plane stretching of atoms in the outer atomic layer leads to $A_{1g}^1$ mode. $E_g$ Raman active modes arise due to symmetric in-plane bending and shearing of atoms in the outer two layers, whereas in $A_{1g}$ mode, atoms vibrate in opposite directions and do result in a shorter displacement of atoms thus producing higher phonon frequency. In Raman active $E_g$ mode, atoms vibrate in the same direction and results in greater atomic displacement of atoms producing lower phonon frequency. The schematics of these vibrations are drawn in Fig. 4(a). A group theoretic analysis has been performed in VIBRATE! Software [35] to get the irreducible representations of the as-grown crystals and determine the Raman active and IR active modes. Using the CIF file generated, the irreducible phonon modes (Raman active, IR active and optically silent modes) are given by:

$$\Gamma_{irr} = [3A_{1g} \oplus 7A_{2g} \oplus 10E_g]_{Raman} \oplus [3A_{2u} \oplus 9E_u]_{IR} \oplus [6A_{1u}]_{Silent} \quad (1)$$

The observed Raman modes agree with the group theory analysed Raman active modes. Fig. 4(b) illustrates the observed Raman spectrum for the synthesized single crystal of BiSbTe$_3$. The $E_g^1$ mode, which occurs below 50 cm$^{-1}$ [36, 37] could not be detected in our measurements due to instrumental limitations. All the other three anticipated Raman active modes are observed as marked in Fig. 4(b). Observed peaks are fitted, analyzed, and identified through Lorentz deconvolution as shown in Fig. 4(b), which reveal that the peaks are broadened or have shoulder peaks. The broadening of peaks is associated with formation of two bonds Te-Bi-Te and Te-Sb-Te. The observed Raman modes and corresponding frequencies obtained by Lorentz deconvolution of obtained peaks are summarized in Table 2. The Raman active $A_{1g}^1$, $E_g^2$ and $A_{1g}^2$ modes vibrational frequencies for studied BiSbTe$_3$ crystals lie in between the parent TIs Bi$_2$Te$_3$ and Sb$_2$Te$_3$ [36-38]. The observed broadened/double Raman peaks for $A_{1g}^1$, $E_g^2$ and $A_{1g}^2$ modes are at frequencies (64.42 ± 0.076 cm$^{-1}$, 73.61 ± 0.646 cm$^{-1}$), (102.39 ± 0.778 cm$^{-1}$, 107.11 ± 0.487 cm$^{-1}$) and (151.03 ± 1.045 cm$^{-1}$, 155.51 ± 0.360 cm$^{-1}$), respectively.

In order to study the electrical transport properties of as-synthesized BiSbTe$_3$ crystal, the four-probe resistance has been measured in the temperature range 2-200 K under zero magnetic field and the same is depicted in Fig. 5(a). The R(T) curve exhibits typical metallic behavior, as the resistance decreases with the lowering of temperature. The R(T) curve behavior is explained by the equation below

$$R_{xx}(T) = R_{xx}(0) + \beta \exp[-\frac{\theta}{T}] + \gamma T^2 \qquad (1)$$

following the work by P. Mal et al [39]. Here, $R_{xx}(0)$ is residual resistance, quadratic term represents electron-electron (*e-e*) interaction and the exponential term accounts for the electron-phonon (*e-ph*) interaction with $\theta$ being the Debye temperature. The R(T) curve is fitted with above equation as shown in Fig. 5(a) and the resulting fitted parameters are $R_{xx}(0)$ = 1.474 mΩ, β = 1.19 mΩ, θ = 82.64 K and γ = 7.76 × $10^{-8}$ Ω $K^{-2}$. A lower θ implies the effective *e-ph* interaction being present in mixed TI at low temperatures in comparison to pure TI. An estimate shows that the strength of *e-ph* interaction equivalent to *e-e* interaction is at $T^*$~25 K. Above $T^*$, the *e-ph* interactions dominate, while below $T^*$, the *e-e* interactions dominate. Another characteristic of resistance of metallic materials is the higher residual resistance ratio, given by $R_{200 K}/R_{0 K}$, which is 3.64. It endorses the quality and metallicity of the as-grown sample $BiSbTe_3$. In comparison to the parent TI compounds, $BiSbTe_3$ has lower metallicity as illustrated in Fig 5(b). For easy comparison, the relative longitudinal resistance $R/R_{50 K}$ is plotted for parent TIs $Bi_2Te_3$ [40], $Sb_2Te_3$[41] and the mixed TI $BiSbTe_3$ under investigation. It shows the higher relative resistance for $BiSbTe_3$ in contrast to other parent TIs. $BiSbTe_3$ is a ternary alloy, which has more structural disorders, defects and scattering centers as compared to the other two binary compounds. These factors result in reduced carrier mobility, which leads to increased relative longitudinal resistivity of $BiSbTe_3$. At low temperatures (<25 K), the *e-ph* interaction in $Bi_2Te_3$ and $Sb_2Te_3$ is quite low as they have higher Debye temperatures i.e., 165 K [42] and 160 K [43], respectively, which is almost two times to that of $BiSbTe_3$. Thus, the higher relative resistance in $BiSbTe_3$ is due to effective *e-ph* scattering present down to low temperatures ($T^*$~25 K). Since the *e-ph* interaction has critical role in the crystal under investigation, explicit *e-ph* coupling constant calculations are warranted. For such calculations, the heat capacity measurement, temperature-dependent or laser power-dependent Raman spectra measurement [44] is suggested for future studies.

The MR% behavior of $BiSbTe_3$ as a function of the magnetic field (H) and temperature (T) is depicted in Fig. 6(a). The MR% shows a symmetric, parabolic dependence on the magnetic field, consistent with the behavior of topological materials exhibiting non-saturating MR [45]. At lower temperatures, such as 2 K, the MR reaches a maximum value of approximately 150% at ±14 Tesla, indicating the dominance of coherent electron transport and topological surface states. As the temperature increases, the MR decreases, with values around 30% at 200 K, highlighting the suppression of quantum effects and increased phonon scattering. This temperature-dependent trend suggests the interplay of bulk and surface contributions, with the surface states being more prominent at low temperatures. The material's linear, non-saturating MR and its significant dependence on temperature make it a promising candidate for topological applications. In contrast to the parent TIs, where the observed MR at 2 K is ~450% for $Bi_2Te_3$ at 5 K and 14 Tesla [40] and ~550% for $Sb_2Te_3$ at 2 K and 12 Tesla [41], the same for $BiSbTe_3$ is suppressed to 150%. The shape of MR curve for mixed TIs also changes from v-type weak anti-localization (WAL) character to U-type in comparison to pure TIs in low field-temperature regime. Such change in MR shape indicates the increased contributions coming from the Bulk if we compare the mixed TI with pure TI. This is also supported by HLN analysis of magnetoconductivity where we find extra scattering contributions including *e-ph* scattering, bulk and surface state coupling and spin-orbit scattering. The reduced MR% in mixed TIs is associated with the presence of *e-ph* coupling effect in $BiSbTe_3$ at low temperatures leads to an increased residual resistance ($R_{xx}(0)$) resulting in reduced MR%.

Further, to investigate the topological nature and contribution in conductivity through different channels, the magneto-conductivity of $BiSbTe_3$ has been analyzed by HLN model. According to the HLN model, for the topological insulator and 2D systems the quantum correction to conductivity modeled by equation[24, 25]:

$$\Delta\sigma(H) = -\frac{\alpha e^2}{\pi h}\left[\ln\left(\frac{B_\varphi}{H}\right) - \Psi\left(\frac{1}{2} + \frac{B_\varphi}{H}\right)\right] \quad (2)$$

where, $B_\varphi = \frac{h}{8e\pi L_\varphi^2}$ is the characteristic field, $L_\varphi$ is phase coherence length, $\psi$ is digamma function, $h$ is Plank's constant and $e$ is the electronic charge. The experimentally observed magneto-conductivity is fitted by the above equation with free parameters $L_\varphi$ and $\alpha$. The polarity of $\alpha$ describes presence of WAL for $\alpha < 0$ and weak localization (WL) for $\alpha > 0$[46]. In TIs, the magneto-conductivity observed in low field regime for WAL effect is described by HLN equation[47, 48]. Fig. 6(b) shows the HLN model fit to low field (±1 Tesla) magneto-conductivity data. The fitted parameters are summarized in Table 3. The negative value of $\alpha$ implies the WAL effect present in the $BiSbTe_3$. The $L_\varphi$ first decreases from 10 K to 50 K, but increases for higher temperatures, which is contrary to the expected. The $L_\varphi$ should decrease with increasing temperature as the *e-ph* scattering become more prominent at higher temperature. Therefore, a modified HNL model [49] $\Delta\sigma(H) = HLN + \beta H^2 + \gamma H$ has been employed to fit the magneto-conductivity in the entire range ±14 Tesla as shown in Fig. 6(c). Here, $\beta$ is the coefficient of the quadratic term that accounts for the elastic scattering and spin–orbit scattering (quantum scattering) in the topological insulator at higher magnetic fields. The linear term with coefficient $\gamma$ represents the bulk contribution in overall conductivity. The fitted parameters are summarized in Table 3. The negative α values imply the presence of WAL effect. Ideally -0.5 value of α account for single conduction channel [50], but here a smaller value ~ -0.1 indicates single channel conduction along with other contributions including *e-ph* scattering, bulk and surface state coupling and spin-orbit scattering. The phase coherence length($L_\phi$) decreases with the increasing temperature as expected and observed in other TIs[51, 52]. The temperature variation of α parameter and ($L_\phi$) is shown in Fig. 6(d). For higher temperatures, coefficient of both quadratic and linear terms decreases, while at 200 K, $\beta$ vanishes which exhibits the decaying quantum scattering effect. The linear term γ remains finite, indicating the contribution to conduction through bulk states. Overall, the HLN analysis reveals that in low temperature regime, the conduction mechanism involves surface and bulk state conduction and quantum scattering effects.

**Conclusion and outlook**

Summarily, we successfully synthesized the $BiSbTe_3$ mixed TI crystals by quicker melt-growth method by shortening the heat-treatment time. XRD measurement on single crystalline flakes confirms the (00$3n$) unidirectional growth of single crystalline mixed TI and the Rietveld refinement of powder XRD asserts the single phase sample without any foreign impurity. The layered growth and planar surface morphology of the as-grown sample has been analyzed and confirmed by FESEM measurement. The Raman spectra exhibits the characteristic peaks of vibrational modes present in the grown sample with broadened peaks elucidating the formation of two types of bonds viz. Te-Bi-Te and Te-Sb-Te. The zero field RT measurement of $BiSbTe_3$ show the typical semi-metallic behaviour with effective *e-ph* interactions being present at low temperatures. This explains the suppressed MR% present in the as-grown $BiSbTe_3$ crystals with respect to the high MR exhibited by pure TIs $Bi_2Te_3$ and $Sb_2Te_3$. The HLN analysis of magnto-conductivity including quadratic and linear correction reveal presence of WAL effect. In low temperature regime, it also indicates that the conduction mechanism in mixed TI is governed by a combination of surface and bulk state conduction along with the quantum scattering effects. The key results emanating from the current work are summarised in the Table 4.

As the transport property analysis shows the significant role of the electron-phonon interaction in low temperature regime leading to suppression in MR. Electron-phonon interaction at low temperatures needs to be quantified by other methods such as temperature/power dependent Raman spectroscopy and will be carried over in the future work. Also, in the present work, we report much faster synthesis of mixed TI. It becomes interesting to see, using similar time saving methods, how the synthesis of other mixed TIs or pure TI turns out and what difference we get from the conventional longer synthesis protocols. This paves the way for future research work.

## Table-1

Parameters obtained from Rietveld refinement:

| Cell Parameters | Refinement Parameters |
|---|---|
| Cell type: Rhombohedral | $\chi^2$=2.89 |
| Space Group: R -3 m | $R_p$=10.5 |
| Lattice parameters: a=b=4.329(4)Å | $R_{wp}$=13.3 |
| & c=30.438(2)Å | $R_{exp}$=7.79 |
| α=β=90° & γ=120° | |
| Cell volume: 380.197Å$^3$ | |
| Density: 21.774g/cm$^3$ | |
| Atomic co-ordinates: | |
| Bi (0,0,0.4037(6)) | |
| Sb (0,0, 0.4037(6)) | |
| Te1 (0,0,0) | |
| Te2 (0,0,0.2033(5)) | |

## Table-2

Raman mode frequencies of as-synthesized BiSbTe$_3$ and other isomorphic TIs

| Raman Mode | $A_{1g}^1 (cm^{-1})$ | $A_{1g}^2 (cm^{-1})$ | $E_g^2 (cm^{-1})$ |
|---|---|---|---|
| BiSbTe$_3$ | 64.42 ± 0.076 | 151.03 ± 1.045 | 102.39 ± 0.778 |
| (Double peaks) | 73.61 ± 0.646 | 155.51 ± 0.360 | 107.11 ± 0.487 |
| Bi$_2$Te$_3$ [38] | 63 | 134 | 103 |
| Sb$_2$Te$_3$ [41] | 69 | 165 | 112 |

## Table-3

HLN and modified HLN fitted parameters for BiSbTe$_3$ single crystal

| Temperature(K) | Low field(±1T) HLN fit | | Modified HLN fit (±14T) | | | |
|---|---|---|---|---|---|---|
| | α | L$_\phi$ (nm) | α | L$_\phi$ (nm) | β(mT$^{-2}$e$^2$/h) | γ(T$^{-1}$e$^2$/h) |
| 10 | -0.09021 | 27.23451 | -0.09275 | 42.25633 | 0.8168 | -0.03067 |
| 50 | -0.1085 | 0.3691 | -0.11667 | 45.91786 | 0.5315 | -0.02272 |
| 100 | -0.8195 | 6.49042 | -0.06136 | 32.19649 | 0.1631 | -0.01003 |
| 200 | -0.01585 | 14.44712 | -0.01358 | 30.33907 | -4.9×10$^{-5}$ | -0.00203 |

## Table-4

Key outcomes of the present work* in comparison to parent TIs

| Key Parameters | Bi$_2$Te$_3$ | Sb$_2$Te$_3$ | BiSbTe$_3$* | Significance |
|---|---|---|---|---|
| Total Growth Time | ~8days [40] | ~10 days [41] | ~3 days | Less time = Larger scale production & lower energy costs. |
| Dominant Scattering | e-ph (weak below 25 K)[40] | e-ph (weak below 25 K)[41] | e-ph > e-e (above 25 K) | Stronger e-ph in BiSbTe$_3$ |
| Debye Temperature (θ) | 165 K[42] | 160 K[43] | 82.64 K | Lower θ Enhanced e-ph at low T |

**Figure Captions**

**Figure 1:** The heat treatment diagram followed for the synthesis of BiSbTe$_3$ single crystals. The inset shows the silvery shiny chunk of obtained crystals.

**Figure 2: (a)** XRD pattern taken on a thin crystal flake of synthesized BiSbTe$_3$ single crystal. **(b)** XRD data taken on powdered sample and peaks analysed by Rietveld refinement of PXRD pattern of grown mixed TI. **(c)** The unit cell of the single crystalline BiSbTe$_3$ generated by VESTA.

**Figure 3: (a)** FESEM image of surface morphology of BiSbTe$_3$ single crystal showing layered growth. **(b)** FESEM image taken on single plane with no color contrast confirm single phase of the grown crystal. **(c)-(e)** Depicts the elemental mapping of the sample exhibiting the homogenious growth of the sample. **(f)** EDAX spectra of synthesized BiSbTe$_3$ single crystal, in which the inset table shows the elemental composition of constituent elements of BiSbTe$_3$.

**Figure 4:(a)** The schematics of different vibrational modes present in the grown BiSbTe$_3$ crystal. **(b)** Room temperature Raman spectra of BiSbTe$_3$ crystal and observed peaks analysed and fitted by Lorentz deconvolution.

**Figure 5: (a)** Temperature-dependent longitudinal resistance of BiSbTe$_3$ crystal at zero field with theoretical fit. **(b)** Relative longitudinal resistance R/R$_{50K}$ for Bi$_2$Te$_3$, Sb$_2$Te$_3$ and BiSbTe$_3$ showing higher relative residual resistance for grown mixed TI.

**Figure 6: (a)** The MR% vs. H plot at 2K, 10K, 50K, 100K and 200K for BiSbTe$_3$. **(b)** Low field (±1T) fitting of magneto-conductivity data for BiSbTe$_3$ by HLN equation **(c)** Plot of magneto-conductivity data for BiSbTe$_3$ and fitted by modified HLN equation in entire field range ±14T. **(d)** Temperature variation of modified HLN fitted parameters.

**Acknowledgment**

The authors would like to thank the Director of the National Physical Laboratory, New Delhi for his keen interest. The authors acknowledge Dr. J. Tawale and Ms. Sweta for FESEM/EDAX and Raman spectroscopy measurements, respectively.

**Data Availability Statement:**

All the data associated with the MS will be made available on reasonable request.

**Conflict of Interest Statement and Declaration:**

The authors have no conflict of interest. The authors declare that they have no known competing financial interests or personal relationships that could have appeared to influence the work reported in this paper.

**Supplementary information:** Not Applicable

**Ethical approval:** Not Applicable

**Figures**
Fig.1

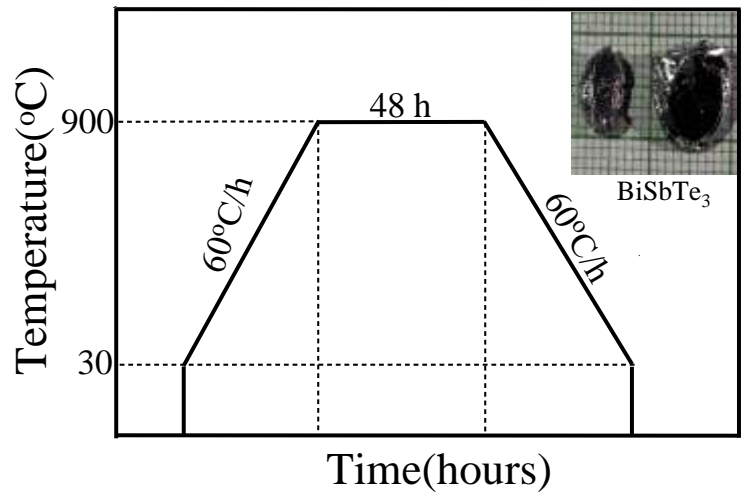

Fig.2

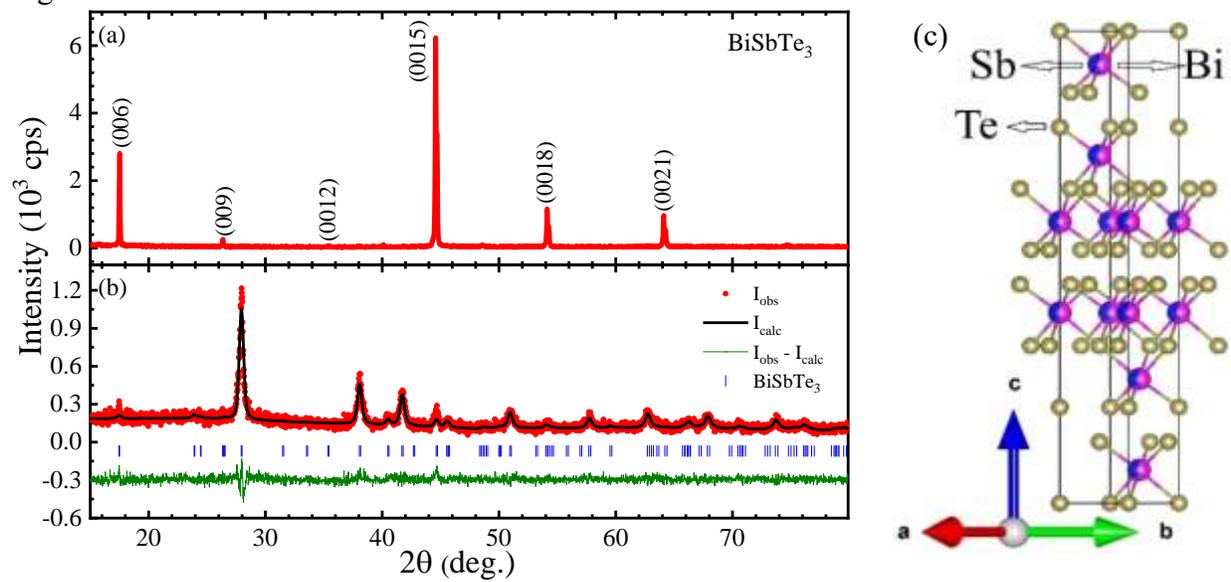

Fig.3

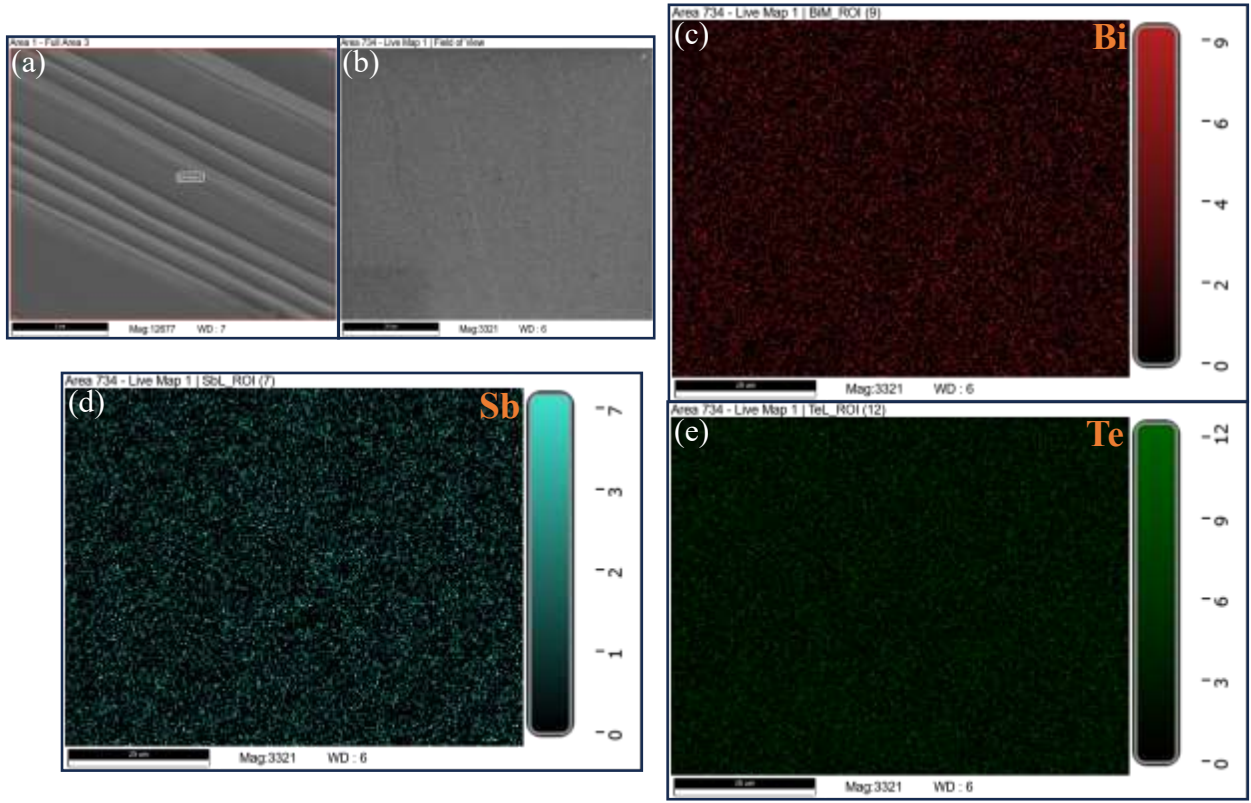

(f)

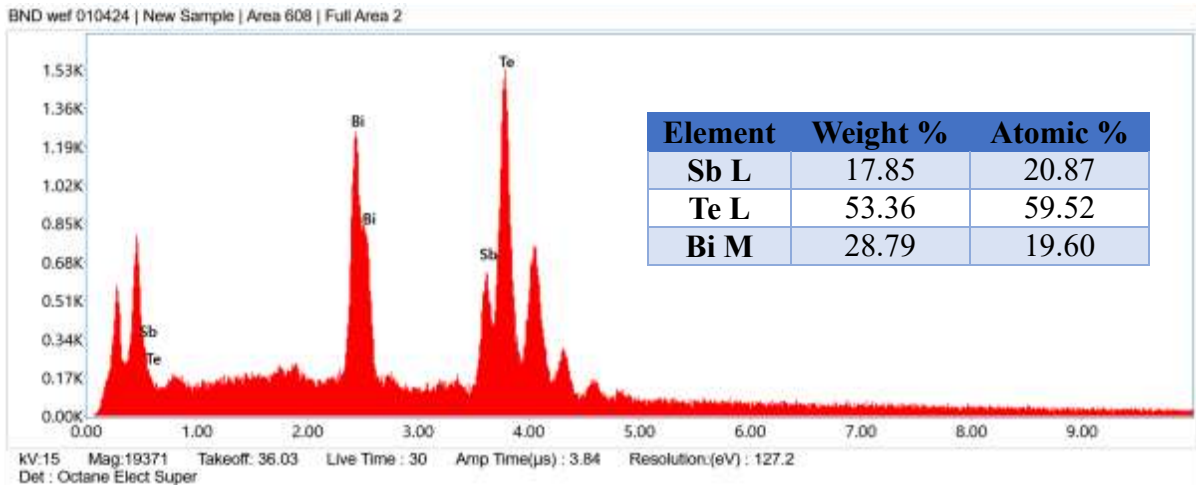

| Element | Weight % | Atomic % |
|---------|----------|----------|
| Sb L | 17.85 | 20.87 |
| Te L | 53.36 | 59.52 |
| Bi M | 28.79 | 19.60 |

Fig. 4
(a)

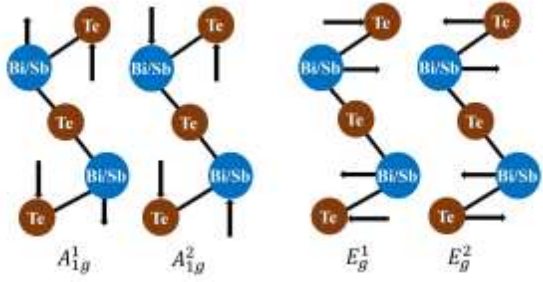

(b)

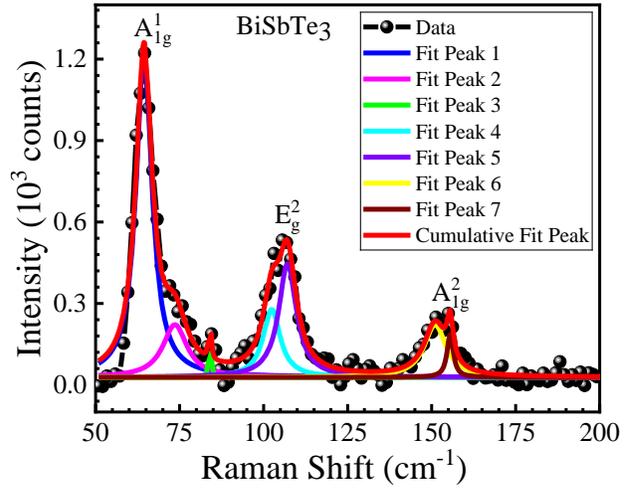

Fig. 5
(a)

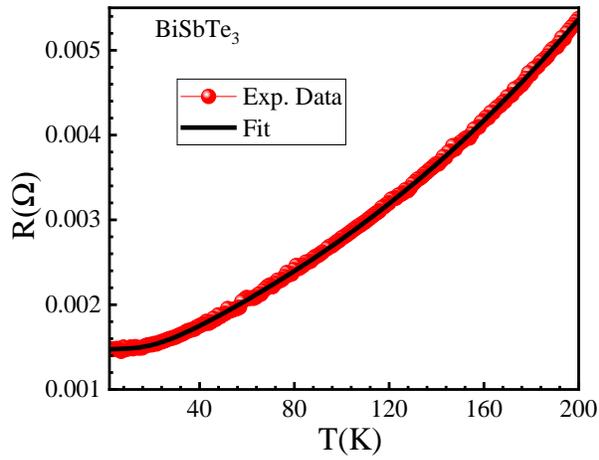

(b)

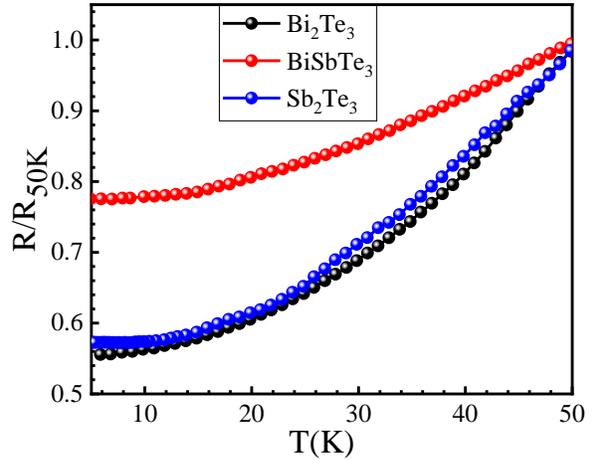

Fig. 6

(a)

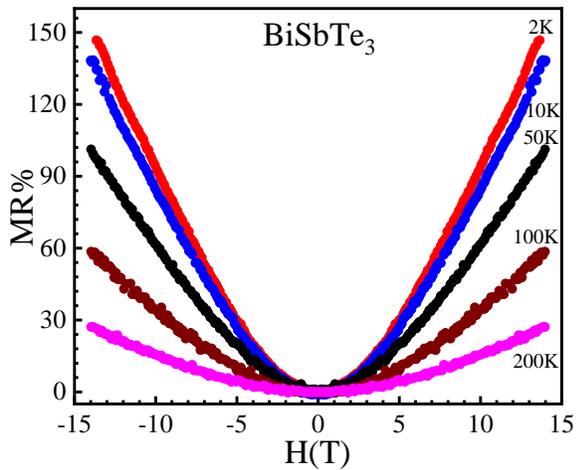

(b)

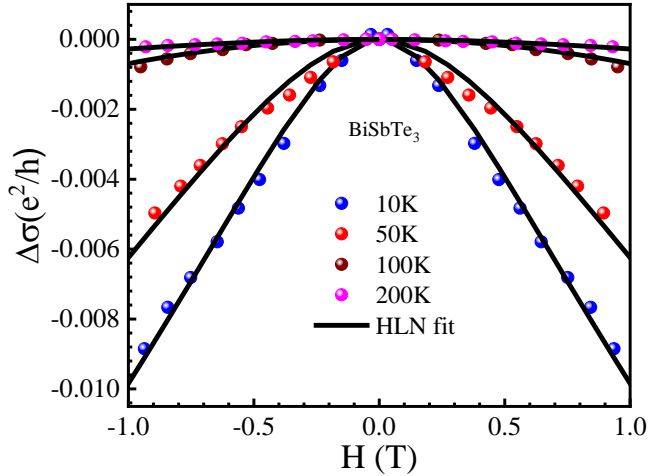

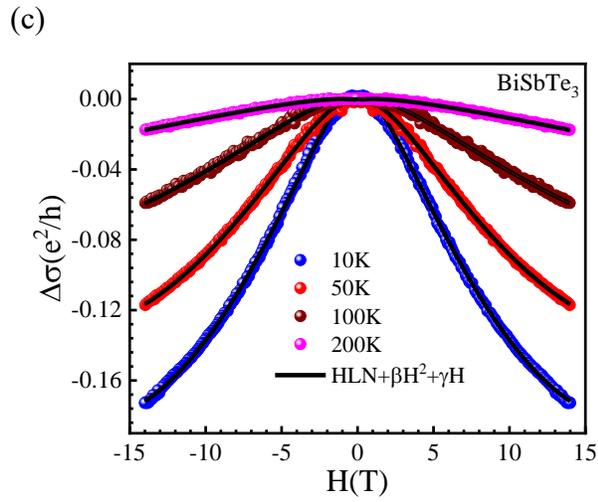
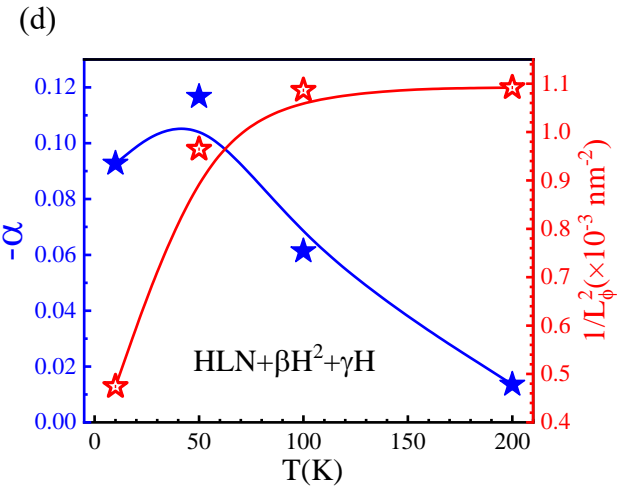

**Graphical Abstract:**

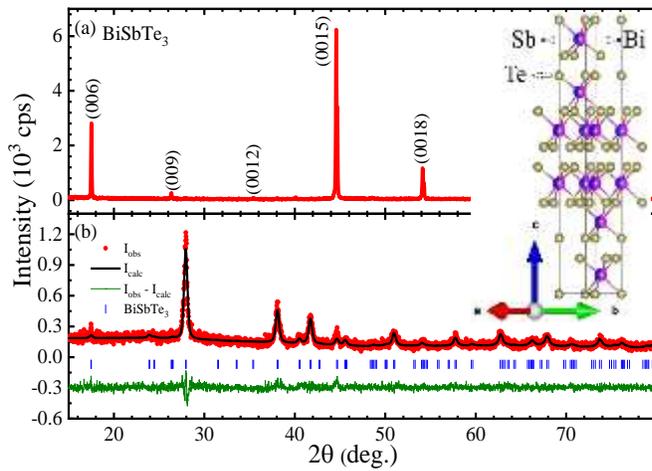

(i) Powder and single crystal XRD with unit cell

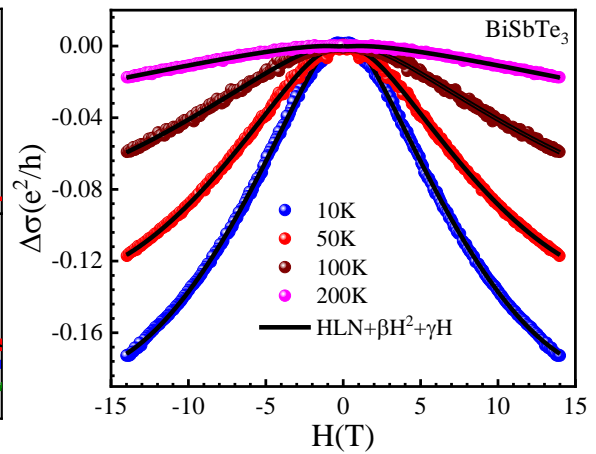

(ii) Magnetoconductivity fitted by modified HLN equation